# Circular Polarization and Quantum Spin:
# A Unified Real-Space Picture of Photons and Electrons

Alan M. Kadin*

Princeton Junction, NJ 08550 USA

August 7, 2005

**Abstract:**

A classical circularly polarized electromagnetic wave carries angular momentum, and represents the classical limit of a photon, which carries quantized spin. It is shown that a very similar picture of a circularly polarized coherent wave can account for both the spin of an electron and its quantum wave function, in a Lorentz-invariant fashion. The photon-electron interaction occurs through the usual electromagnetic potentials, modulating the frequency and wavevector (energy and momentum) of this rotating spin field. Other quantum particles can also be represented either as rotating spin fields, or as composites of such fields. Taken together, this picture suggests an alternative interpretation of quantum mechanics based solely on coherent wave packets, with no point particles present.

*E-mail: amkadin@alumni.princeton.edu

## I. Introduction

Maxwell's equation have long proven to be extremely productive and insightful, in ways that Maxwell and his contemporaries could not have imagined. In fact, these equations contain within them the seeds of special relativity, providing the motivation for Einstein to develop his theories. But quantum mechanics has always seemed to stand quite apart from Maxwell's equations, despite the central role played by light and optical phenomena in the early historical development of quantum theory, from black-body radiation and atomic spectra to the photoelectric effect. The key development, of course, was the photon, with quantized energy $E = \hbar\omega$.

The present paper proposes that contrary to conventional wisdom, the quantum nature of the photon flows quite naturally from its classical electromagnetic properties, based on a simple picture of a coherent, circularly polarized wave packet, with its rotating electric field vector which carries angular momentum. The only additional assumption is quantization of spin $\hbar$. It is further shown that a very similar picture of a vector field rotating coherently at $\omega = mc^2/\hbar$ can also account directly for both the spin $\hbar/2$ and the quantum wavefunction of massive quantum particles such as the electron [1,2]. The deBroglie wavelength falls out naturally as a consequence of Lorentz invariance, and this rotating vector field maps onto the complex scalar field of the Schrödinger equation. Within this picture, there are no point particles; there are only distributed, coherent wave packets with quantized total spin.

Quantized spin provides the bridge from classical waves to quantum waves. Historically, spin was regarded as an incidental feature of some quantum particles, separate from the fundamental quantum nature. On the contrary, it is seen here that a rotating spin field provides for the very existence of quantum waves in fundamental quantum particles such as the photon and the electron. The electromagnetic potentials of this photon wave modulate the frequency and wavevector of the electron wave, in a way that is consistent with conservation of energy, momentum, and angular momentum for the corresponding particles. Composite particles derive their quantum properties from the underlying spin fields of their components. This presents a picture of quantum mechanics that seems quite different from the standard statistical interpretation, but appears consistent with the standard quantum equations and results.

**II. Circularly Polarized EM Waves and the Photon**

A transverse wave such as an electromagnetic wave is generally constructed from linearly polarized sine wave components of the form, for example, $\hat{\mathbf{x}}\cos(\omega t - kz)$ for a wave propagating in the *z*-direction, oscillating in the *x*-direction. But one can equally well choose a set of circularly polarized (CP) components, of the form

$$\mathbf{E} = E_0[\hat{\mathbf{x}}\cos(\omega t - kz) \pm \hat{\mathbf{y}}\sin(\omega t - kz)], \tag{1}$$

where the ± corresponds to the two opposite helicities. Any linearly polarized wave, or indeed any general elliptically polarized wave, can be constructed by an appropriate linear combination of circularly polarized waves of both helicities. Note that this CP wave corresponds to a vector of fixed length $E_0$, rotating about the *z*-axis at a fixed $\omega$, so that it is natural to represent in a polar picture as

$$\mathbf{E} = E_0 \angle \theta, \quad \text{where} \quad \theta = \pm(\omega t - kz), \tag{2}$$

where again the ± corresponds to the opposite helicities. This rotation of a real vector about a fixed axis is mathematically equivalent to rotation of a complex scalar in the complex plane, where *x* maps onto the real axis and *y* onto the imaginary axis. Only a CP wave has this unique property, which suggests a close connection with a complex scalar quantum wavefunction.

Such a CP wave may be localized in space in the form of a wave packet, if the amplitude $E_0$ is reduced toward the edges. Fig. 1 shows the helical form of such a wave, attenuating in the direction of motion. It is such a CP wave packet that we would like to identify below as the real-space picture of a photon.

It is well known that an electromagnetic wave carries energy and momentum, associated with the Poynting vector $\mathbf{P} = \mathbf{E} \times \mathbf{H} = (\mathbf{E} \times \mathbf{B})/\mu_0$, and distributed through the wave.

Here **B** and **H** are the usual magnetic vectors and $\mu_0$ is the usual permeability of free space (SI units are used here and throughout the paper). One defines an energy density $\mathcal{E}$ and momentum density $\mathcal{P}$ given by the following expressions:

$$\mathcal{E} = |\mathbf{P}|/c = |\mathbf{E} \times \mathbf{B}|/\mu_0 c = \varepsilon_0 E^2 \tag{3}$$

$$\mathcal{P} = \mathcal{E}/c = (\mathbf{E} \times \mathbf{B})/\mu_0 c^2 = \varepsilon_0 E^2 / c \tag{4}$$

It is perhaps somewhat less well known (but still a standard result [3,4,5]) that for a CP wave, one may also define a spin angular momentum density $\mathcal{S}$, which is given by

$$\mathcal{S} = |\mathbf{E} \times \mathbf{A}|/\mu_0 c^2 = \varepsilon_0 E^2 / \omega, \tag{5}$$

where **A** is the usual vector potential given by $\mathbf{B} = \nabla \times \mathbf{A}$. This spin density $\mathcal{S}$ has the following relations to $\mathcal{E}$ and $\mathcal{P}$:

$$\mathcal{E} = \mathcal{S}\omega, \quad \mathcal{P} = \mathcal{S}k, \tag{6}$$

where as usual for an EM wave in free space, $k = \omega/c$. Note the similarity in form between Eq. (6) and the Einstein-de Broglie relations $E = \hbar\omega$ and $p = \hbar k$; the significance of this will be more evident below.

It is also well known that electromagnetic waves in free space are Lorentz covariant; they may be red-shifted or blue-shifted arbitrarily, but the dispersion relation $\omega = kc$ is retained. If one Lorentz-transforms a CP wave with its spin axis and velocity $c$ in the $z$-direction, it remains a CP wave with speed $c$ and spin parallel to the $z$-direction, even if the transformation is in the $x$- or $y$-directions.

With the advent of the quantum theory, it became evident that EM waves are quantized into units known as photons. Each photon has quantized angular momentum along the direction of motion of $\pm\hbar$, and it is accepted that the CP wave represents the classical limit of a photon with energy $E = \hbar\omega$. It is suggested here that this picture be taken quite literally, that the quantization of spin defines the photon. Within this picture, the photon is not a point particle, but a coherent CP wave packet in a region of space with a total spin angular momentum of $\pm\hbar$. By integrating Eq. (6) over the wave packet, one obtains the Einstein-deBroglie relations $E = \hbar\omega$ and $p = \hbar k$ as corollaries. It also follows from the dispersion relation that $E = pc$, in accordance with a massless "particle" moving at the speed of light. One can also rewrite the phase angle of the spinning field vector in the form

$$\theta = \pm (Et - \mathbf{p} \cdot \mathbf{r})/\hbar, \qquad (7)$$

which is now clearly Lorentz-invariant, as the inner product of two 4-vectors, ($E/c$, $\mathbf{p}$) and ($ct$, $\mathbf{r}$). These relationships are summarized in Table I.

A single photon would correspond to a coherent circularly polarized wave packet at frequency $\omega$, on the scale of (at least) the order of the wavelength $\lambda = 2\pi c/\omega$, with an intensity distribution such that the total spin is $\hbar$ and the total energy is $\hbar\omega$. For relatively low frequencies, such a photon is not small. For example, consider a radio-frequency photon at $f = 1$ MHz, with wavelength $\lambda = 300$ m, volume $\lambda^3$, and energy ~4 neV. A photon on this scale would correspond to an electric field of order $10^{-12}$ V/m. In contrast, for an x-ray photon at energy 1 keV, with wavelength 1.2 nm, a photon on this scale yields $E_0 \sim 10^{+11}$ V/m. This photon is much smaller, but it is still a distributed wave packet rather than a point particle.

It is common to regard a photon as a traveling wave, but one may also have a standing-wave photon, as in a resonator. In this case, one would have a superposition of wave components in different directions, leading in the usual way to spatial nodes where the amplitude goes to zero. For example, if one has boundary conditions with mirrors on the two ends of a length L where the electric field goes to zero, one has a coherent photon state only for conditions where an integral number of half-wavelengths can fit in L:

$$\mathbf{E_n} = E_0 [\hat{\mathbf{x}} \cos(\omega t) \pm \hat{\mathbf{y}} \sin(\omega t)] \sin(k_n z), \qquad (8)$$

where $k_n = n\pi/L$, so that there are a set of quantized energy levels $E_n$ given by

$$E_n = \hbar\omega = \hbar c n\pi/L, \qquad (9)$$

for either helicity. The normalization amplitude $E_0$ is determined by integrating the energy density $\mathcal{E}$ over the volume of the photon state and setting equal to the energy $E_n$, or equivalently by integrating $\mathcal{S}$ and setting equal to $\hbar$.

### III. Circularly Polarized *Electron* Field

Now consider a similar derivation with a vector field that represents not a massless photon, but rather a massive particle such as the electron. Let us assume a coherent CP field similar to that in Eq. (1) or in Fig. 1:

$$\mathbf{\Psi} = \Psi_0[\hat{\mathbf{x}}\cos(\omega t - \mathbf{k}\cdot\mathbf{r}) \pm \hat{\mathbf{y}}\sin(\omega t - \mathbf{k}\cdot\mathbf{r})] = \Psi_0 \angle \theta \quad , \tag{10}$$

and for now let us also assume that **k** is in the *z*-direction, parallel to the spin axis. The use of the sympol $\mathbf{\Psi}$ is deliberate; this two-dimensional rotating vector will turn out to map directly onto the complex quantum wavefunction. We can think of this as the *Electron* field in contrast to the *Electric* field of the photon. But for now let us assume only that the various densities ($\mathcal{E}$, $\mathcal{P}$, and $\mathcal{S}$) are proportional to the intensity $\Psi_0^2$ (as for the photon), together with a slightly modified equation relating these densities:

$$\mathcal{E} = 2\mathcal{S}\omega, \quad \mathcal{P} = 2\mathcal{S}k \tag{11}$$

Why the extra factor of 2? That is necessary to recover the Einstein-deBroglie relations, if we take the total spin of the electron wave packet to be $S = \hbar/2$. This factor of 2 is not compatible with Maxwell's equations, but remember that here we want to represent a massive electron moving with $v<c$, rather than a massless particle at $v=c$. It is not hard to construct a simple semiclassical model that reproduces Eq. (11) (see ref. [1]), but these relations should be true more generally than any particular model. Also, Eq. (11) appears to be consistent with the relativistic Dirac equation for the electron [4].

For such a spin-1/2 particle, the Lorentz-invariant rotation angle $\theta$ can again be written in the form:

$$\theta = (Et - \mathbf{p}\cdot\mathbf{r})/\hbar, \tag{12}$$

where this is Lorentz-invariant if and only if $E$ and **p** are the proper covariant forms for a massive particle:

$$E = \gamma mc^2, \quad \mathbf{p} = \gamma m\mathbf{v}, \text{ for } m>0 \text{ and } v<c, \tag{13}$$

where $m$ is the rest mass and $\gamma = 1/\sqrt{(1-v^2/c^2)}$. In the nonrelativistic limit $v \ll c$, this becomes

$$\omega \approx (mc^2 + mv^2/2)/\hbar \text{ and } k \approx mv/\hbar. \tag{14}$$

Note that this yields the expected de Broglie wavelength $\lambda = h/mv$ by Lorentz invariance only if the rest energy of the electron is included in $\omega$; the zero of energy is not arbitrary here. Also, we have defined this rotating spin wave in terms of a particle with mass $m$, but it could just as easily have been defined locally by a rest-mass density $\mathcal{M} = \mathcal{E}/c^2 \propto \Psi_0^2(\mathbf{r})$. The "particle properties" such as mass, energy, and momentum all derive from the local properties of the wave, integrated over a volume defined by quantized spin.

Unlike the case of the photon, one can Lorentz-transform such a wave packet into its rest frame with $v=0$. In this case, the rotation angle $\theta$ is uniform across the wave packet, rotating at a minimum frequency $\omega = mc^2/\hbar$. For the electron with $mc^2 = 511$ keV, this corresponds to a frequency $f = \omega/2\pi \sim 10^{20}$ Hz, corresponding to gamma-ray frequencies for photons. This rotation is intrinsic, associated with the spin, and cannot be slowed or stopped.

Furthermore, once this has been transformed to the rest state, it is no longer a transverse wave. In fact, it can then be transformed in any direction, not limited to the spin axis. For example, one can construct a wave spinning around the $z$-axis but moving in the $x$-direction. The motion of this vector is no longer a simple helix as in Fig. 1, but would exhibit more general cycloidal motion. In effect, the spin and orbital motions are essentially decoupled. Again, this is possible (in contrast to the photon case) only because this wave is moving at a speed $v<c$.

As before for the photon, boundary conditions can give rise to standing waves which maintain the rotating spin field while giving spatial nodes. For a box of length $L$, Eq. (8) for the eigenfunction becomes

$$\mathbf{\Psi_n} = \Psi_0[\hat{\mathbf{x}}\cos(\omega t) \pm \hat{\mathbf{y}}\sin(\omega t)]\sin(k_n z), \tag{15}$$

where $k_n = n\pi/L$ now yields quantized energies, for $v \ll c$,

$$E_n = \hbar\omega = mc^2 + \hbar^2 k^2/2m = mc^2 + (\hbar n\pi)^2/2mL^2, \tag{16}$$

where of course the energies are the same for either helicity.

These are the same set of energies that one would obtain from the conventional nonrelativistic Schrödinger equation, apart from the $mc^2$ offset in energies. One can formalize this equivalence by defining a complex wavefunction $\Psi_c$ by

$$\Psi_c = \Psi_x \mp \Psi_y = \Psi_0 \exp(\mp\theta). \tag{17}$$

Then

$$i\hbar\partial\Psi_c/\partial t = \hbar\omega\Psi_c = \left(mc^2 + \hbar^2 k^2/2m\right)\Psi_c = (mc^2 - \hbar^2\nabla^2/2m)\Psi_c, \tag{18}$$

which confirms the mapping to the Schrödinger equation, with the rest-energy offset. More generally, there would be a potential energy term included in the energy in both the standard and the rotating spin field formulations.

The relations for the electron rotating spin field are summarized in Table II, for direct comparison with the photon in Table I. These are qualitatively quite similar; both are coherently rotating vector fields with quantized spin. The photon has spin-1, $m=0$, and $v=c$, while the electron has spin-1/2, $m>0$, and $v<c$. Furthermore, the electronic charge has not even been mentioned, so that this massive spin field could equally well represent any spin-1/2 lepton or quark. Even the neutrino is now believed to have a finite mass, and so might also fit into the electron field picture.

The major implications of this overall picture should now be clear. The complex quantum wavefunction is simply a mathematical representation of a rotating coherent vector spin field having a quantized total spin. It is this spin quantization that couples local wave properties ($\omega$ and **k**) to global particle properties (E and **p**). Intrinsic spin is the essential basis for quantum mechanics.

### IV.  Discussion

It was proposed above that the quantum wave function for massive fundamental particles is a mathematical representation of rotation at $mc^2/h$. But quantum effects are known to exist for bound composites of multiple fundamental particles, even if they have no net spin. For example, atoms show quantized vibrational and rotational states, and in fact the observation of depressed heat capacities of atomic and solid systems at low temperatures (due to these quantized energy levels) was historically one of the motivations for the development of the quantum theory. Is that consistent with this rotating spin-field picture?

This is indeed consistent, provided that all of the component particles remain in coherent quantum states. This can be illustrated using a standard result [6] that the complex quantum wavefunction $\Psi_{tot}$ of an N-particle state is in many situations the product of the complex wavefunctions $\Psi_n$ of the components:

$$\Psi_{tot} = \prod_{n=1}^{N} \Psi_n = \prod_n [\Psi_{n0} \exp(-iE_n t/\hbar)] = \exp(-iE_{tot} t/\hbar) \left[ \prod_n \Psi_{n0} \right], \quad (19)$$

where $E_{tot} = \Sigma E_n$ is the total energy of the state. If each of these energies includes the rest energy, then the total energy includes the term $Mc^2$ for the total mass $M = \Sigma m_n$ The composite wavefunction is then acting "as if" it represents a vector field rotating at $Mc^2/h$. But in fact, the quantum phenomena are due to the collective interference effects of each of the component waves.

One might initially think that a rapidly rotating vector could be measured directly, but since this is an intrinsic rotation that cannot be slowed, the situation is not quite so obvious. In general, one needs to consider the interaction of a quantum wave with its environment in order to extract a meaningful measurement.

One type of interaction is between an electron and a photon. For a classical point electron in a classical electromagnetic field, this interaction can be expressed in terms of the electromagnetic potentials $V$ and $\mathbf{A}$ (which generate the fields $\mathbf{E}$ and $\mathbf{B}$) which modify the energy $E$ and momentum $\mathbf{p}$ of the particle:

$$E \to E + eV; \quad \mathbf{p} \to \mathbf{p} + e\mathbf{A} \quad (20)$$

This is dependent on the gauge, but a suitable consistent gauge choice can normally be identified. These relations are carried over into quantum mechanics through the Einstein-deBroglie relations, so that

$$\omega \to \omega + eV/\hbar; \quad \mathbf{k} \to \mathbf{k} + e\mathbf{A}/\hbar \quad (21)$$

For a typical electronic transition that involves absorption or emission of a photon, $V$ and $\mathbf{A}$ are oscillatory at a frequency $\omega_{ph}$ and wavevector $\mathbf{k}_{ph}$. One can then regard $V$ and $\mathbf{A}$ as modulating $\omega$ and $\mathbf{k}$, respectively:

$$\omega \to \omega \pm \omega_{ph}; \quad \mathbf{k} \to \mathbf{k} \pm \mathbf{k}_{ph} \tag{22}$$

This is directly analogous to frequency modulation of radio signals, where a properly phased modulation can give rise to either a sum or a difference of frequencies. When one multiplies by $\hbar$ again, this in turn leads to a transition to a final state that conserves energy and momentum. This a standard result [6], but it provides a consistent microscopic foundation for the conservation laws in a picture of distributed rotating spin waves. One can even view the particle-based conservation laws as following from local wave-based interactions, with spin quantization connecting the two.

To be fully consistent, one also needs to consider the back-action effect of the electron on the photon. By analogy, one should be able to construct a set of electronic potentials $V_e$ and $\mathbf{A}_e$, the generators of the rotating electron field $\Psi$. These would add to the photon energy and momentum in the same way that the electromagnetic potentials add to the electron energy and momentum above. This would correspond to appropriate sums and differences of $\omega$ and $\mathbf{k}$ within the local photon wave, assuring energy and momentum conservation in processes such as Compton scattering.

The previous analysis has focused on similarities between photons and electrons. However, a key difference between spin-1/2 fermions such as the electron and spin-1 bosons such as the photon is that the former are subject to the Pauli exclusion principle, whereby only a single electron (of each spin polarity) can be placed in a given eigenstate. Although this is somewhat beyond the scope of the present paper, one may speculate that the same interaction that gives rise to spin quantization may also give rise to the Pauli exclusion principle. For example, one might have a self-interaction energy for fermions that changes sign when the cumulative spin of a coherent field in a region exceeds $\hbar/2$.

Finally, a fully wave-based picture of quantum mechanics is likely to have important implications for quantum measurement theory. For example, a typical single-photon detector in the optical range involves excitation of a localized electron state on a scale that is much smaller than the optical wavelength (or the photon size). The wave picture suggests that the transition takes place via a continuous dynamical wave process, directly analogous to detection of a radio wave with an antenna much smaller than the wavelength. No discontinuous "collapse of the wave function" is necessary. But for the photon detector, there are typically many available electron localized states, and any given photon transfers its energy to just one, rather than splitting the energy among them. The key to this must be the quantization of spin, which assures that the measurement is not complete until the full $\hbar$ of spin has been transferred to a single electronic state. The situation is even more complicated for entangled states of correlated pairs of photons or electrons, which have received attention recently in the context of quantum information theory [7]. But here, too, it may be instructive to consider the picture of extended coherent wave states, rather than the more common particle picture.

## V.    Conclusions

It is shown that a simple picture based on a coherent circularly polarized electromagnetic wave packet can form the basis of a microscopic wave picture of the photon. The full set of quantum relations fall out naturally via Lorentz invariance, with the single assumption of quantization of spin. A very similar picture of a coherently rotating vector field with distributed mass-energy leads to the quantum wavefunction for the electron and other fermions. Conservation of energy and momentum for quantum particles are seen to follow from wave-level interactions modulating rotation frequency and wavevector. It is suggested that all quantum effects may be due to rotating vector fields in fundamental particles with spin, and composites of such particles.

This picture is conceptually quite different from the prevailing Copenhagen interpretation of quantum mechanics, in that there are no ensemble probabilities of particles, just interacting wave packets. But the resulting equations appear to be consistent with standard quantum theory. If that is true, does it really matter which alternative interpretation one uses? In fact, quantum mechanics has a reputation as being fundamentally confusing and full of paradoxes, despite its great success as a calculational tool. One can make an analogy with electromagnetics at the end of the 19$^{th}$ century, where considerable efforts were focused on the paradoxical behavior of the "luminiferous ether" that was believed to carry electromagnetic waves. With the advent of special relativity, all of these problems disappeared, together with the need for the ether itself. It is to be hoped that when quantum mechanics is viewed from an appropriate consistent and complete foundation, many of the present quantum paradoxes will likewise be resolved.

**References**

[1]     A.M. Kadin, "Spin as the Basis for Quantum Mechanics: A New Semiclassical Model for Electron Spin", ArXiv Quantum Physics preprint, available online at http://arxiv.org/abs/quant-ph/0504068 (2005).

[2]     A.M. Kadin, "Quantum Mechanics without Complex Numbers: A Simple Model for the Electron Wavefunction Including Spin", ArXiv Quantum Physics preprint, available online at http://arxiv.org/abs/quant-ph/0502139 (2005).

[3]     J.D. Jackson, *Classical Electrodynamics*, 3$^{rd}$ ed. (John Wiley, New York, 1999), p. 350.

[4]     H.C. Ohanian, *Am. J. Phys.* **54**, 500 (1986).

[5]     K. Gottfried, *Quantum Mechanics: Fundamentals*, (Addison-Wesley, Reading, Mass., 1989), p. 412 [reprint of 1966 Benjamin edition].

[6]     See any of the classic textbooks on quantum mechanics, e.g., E. Merzbacher, *Quantum Mechanics*, 3$^{rd}$ ed. (John Wiley, New York, 1997).

[7]     A. Zeilinger, G. Weihs, T. Jennewein, and M. Aspelmeyer, *Nature* **433**, 230 (2005).

Table I.  Photon in Rotating Spin Field Picture

| Circ. Pol. Electric Field Vector **E** | $\mathbf{E} = E_0[\hat{\mathbf{x}}\cos(\omega t - kz) + \hat{\mathbf{y}}\sin(\omega t - kz)] = E_0 \angle \theta$ |
|---|---|
| Wave equation | Maxwell's Equations |
| Rotating Phase Angle $\theta$ | $\theta = \omega t - kz$ |
| Energy Density $\propto$ Amplitude$^2$ | $\mathcal{E} = \varepsilon_0 E_0^2$ |
| Angular Momentum Density $\mathcal{S}$ | $\mathcal{E} = \mathcal{S}\omega,\ \mathcal{P} = \mathcal{S}k$ |
| Quantization of Spin $S$ | $S = \hbar \Rightarrow E = \hbar\omega,\ p = \hbar k$ |
| Particle Energy-Momentum relation | $E = pc$, for $m=0$ and $v=c$ |
| Lorentz-invariant phase angle | $\theta = (Et - pz)/\hbar$ |
| Effects of Lorentz transformation | Always $v=c$ and $S \parallel p$, no minimum $\omega$ |

Table II.  Electron in Rotating Spin Field Picture

| Circ. Pol. *Electron* Field Vector $\psi$ | $\mathbf{\Psi} = \Psi_0[\hat{\mathbf{x}}\cos(\omega t - \mathbf{k}\cdot\mathbf{r}) + \hat{\mathbf{y}}\sin(\omega t - \mathbf{k}\cdot\mathbf{r})] = \Psi_0 \angle \theta$ |
|---|---|
| Wave Equation | Maps onto Complex Schrödinger Equation |
| Rotating Phase Angle $\theta$ | $\theta = \omega t - \mathbf{k}\cdot\mathbf{r}$ |
| Energy Density $\propto$ Amplitude$^2$ | $\mathcal{E} \propto \Psi_0^2$ |
| Angular Momentum Density $\mathcal{S}$ | $\mathcal{E} = 2\mathcal{S}\omega,\ \mathcal{P} = 2\mathcal{S}k$ |
| Quantization of Spin $S$ | $S = \hbar/2 \Rightarrow E = \hbar\omega,\ p = \hbar k$ |
| Particle Energy-Momentum relation | $E = \gamma mc^2,\ p = \gamma mv$, for $m>0$ and $v<c$ |
| Lorentz-invariant phase angle | $\theta = (Et - \mathbf{p}\cdot\mathbf{r})/\hbar$ |
| Effects of Lorentz transformation | Can have $v=0$ with min. $\omega = mc^2$, no need for $S \parallel p$ |

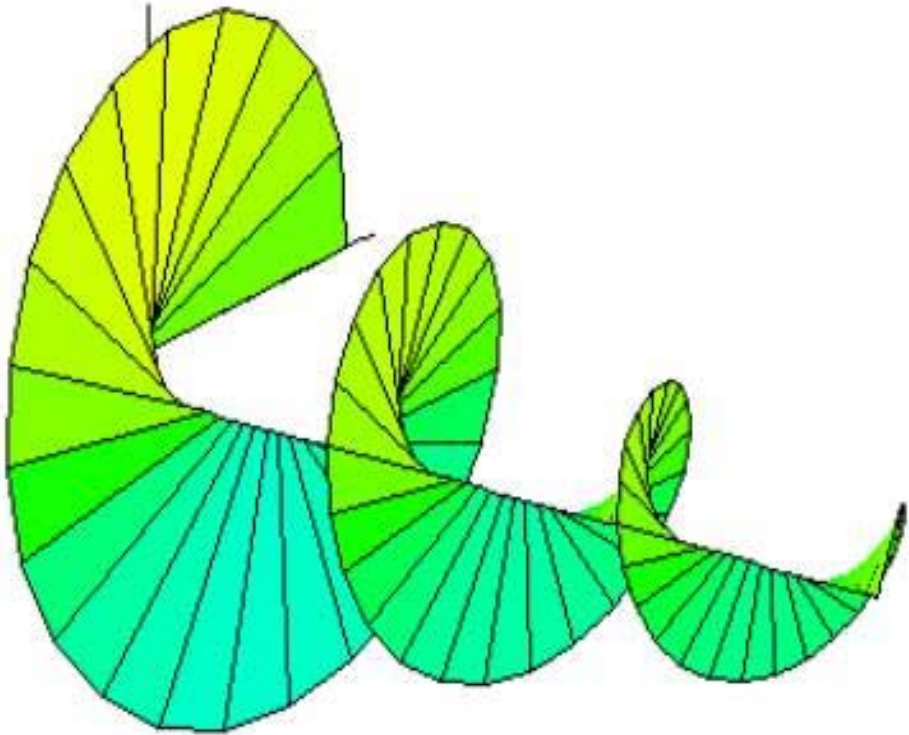

**Fig. 1.** Helically rotating electric field vector corresponding to a circularly polarized (CP) electromagnetic wave propagating with attenuation to the right. A similar picture of a finite wave packet represents a real-space picture of a quantized photon, as well as an electron in the rotating-spin-field picture.